\documentclass[showpacs,preprintnumbers,two column]{revtex4}
\usepackage{graphicx}
\usepackage{dcolumn}
\usepackage{bm}

\begin{document}

\title{Large-N scaling behavior of the quantum fisher information in the
Dicke model }
\author{Yu-Yu Zhang$^{*}$, Xiang-You Chen}
\address{Department of Physics, Chongqing University, Chongqing
400044, P. R.  China
}
\begin{abstract}
Quantum Fisher information (QFI) of the reduced two-atom state is employed
to capture the quantum criticality of the superradiant phase transition in
the Dicke model in the infinite size and finite-$N$ systems respectively.
The analytical expression of the QFI of its ground state is evaluated
explicitly. And finite-size scaling analysis is performed with the large
accessible system size due to the effective bosonic coherent-state
technique. We also investigate the large-size scaling behavior of the scaled
QFI of the reduced $N$-atom state and show the accurate exponent.
\end{abstract}

\maketitle

\section{Introduction}

Quantum Fisher information (QFI), one of the quantum information-based
tools, is a basic concept in quantum estimation theory, which depicts the
theoretical bound for the variance of an estimator~\cite%
{maccone,plenio,brody,holevo,helstrom,wang,yang}. The Fisher information is
the central notion in parameter estimation due to the Cram\'{e}r-Rao
inequality, which sets a basic lower bound to the variance of any unbiased
estimator in terms of the Fisher information~\cite{cover}. Consider the
problem of estimating a unknown parameter $\theta$ from a quantum state $%
\rho(\theta)$. The value of $\theta$ can be estimated from the measurement
results of a proper physical observable. Since the precision of the
estimation is limited by unavoidable measurement errors, the inverse of the
QFI provides the lower bound of the error of the estimation of $\theta$.

Latterly, a new emphasis has emerged in which QFI is related to properties
of interacting many-body systems. This approach is being pursued most
vigorously in connection with quantum phase transition (QPT)~\cite{sachdev},
as it is hoped that the QFI may shed light upon the dramatic effects
occurring in critical systems. QPT and quantum-critical phenomena occur at
zero temperature in many-body quantum systems. A dramatic change of order
parameters exhibits at the critical point, which are induced by the change
of parameters in quantum critical systems. There is much on-going interest
to test the quantum criticality of the QFI in the proximity of phase
transitions~\cite{salvatori,zanardi,venuti,paris}. It is expected to
characterize the singularity of the QPT from quantum estimation perspective
by driving the system toward critical points. In this framework, we consider
the scaling exponents of the QFI at the critical point in the Dicke model~%
\cite{dicke}, which is a well-known quantum collective atoms model.

The Dicke model describes the interaction of $N$ two-level atoms with a
single bosonic mode. The QPT was explored in the Dicke model, exhibiting a
superradiant phase transition in the thermodynamics limit~\cite{Emary,lambert}. 
Although the Dicke model cannot be solved analytically, an
extended bosonic coherent state approach can solve the Dicke model numerical
accurately for large size systems~\cite{chen}. For finite-size atoms Dicke
model has been characterized in terms of entanglement of its ground states~%
\cite{vidal,vidal1,vidal2,liberti,wangchen}, fidelity susceptibility~\cite%
{liu} and the Berry phase~\cite{gang}. However, the quantum criticality in
terms of QFI has not been well analyzed, except preliminarily results for
the QFI of the field mode and $N$-atom state in the ground state~\cite{jin}.
A convincing finite-size scaling behavior of the QFI is still lacking. To
the best of our knowledge, the finite-size studies are limited to numerical
diagonalization in the bosonic Fock state in small-size systems $N\leq35$~%
\cite{emary1,emary2}. Our paper is intended to propose the QFI of the
reduced two-atom state to study the quantum criticality of the QPT in the
Dicke model and the finite-size scaling exponents.

In this paper, we study the QFI of the reduced two-atom state in the Dicke
mode in infinite size and finite-$N$ systems respectively, giving the
accurate finite-size scaling exponents. The paper is organized as follows.
In Sec. II we review the definition of the QFI and its physical signatures
by the parameter estimation theory. In Sec.III we introduce the background
of the Dicke model, and define the QFI of the reduced two-atom state to
capture the quantum criticality in the large-$N$ atoms system and in the
thermodynamics limit. In Sec.IV the large $N$ scaling behavior of both of
the QFI of the two-atom state and $N$-atom state are calculated by the
bosonic coherent-state technique, giving the accurate scaling exponents.
Finally, we summarize our work in Sec.IV.

\section{general formalism for the QFI}

To begin with, we briefly review the parameter estimation theory and the
QFI, which is applied to evaluate bounds of the variance of estimator for a
parameter that we can get from a quantum state. To measure the precision of
the estimator $\theta$, we consider a quantum state $\rho (\theta )$. The
generalized quantum Fisher information (QFI) $F$ is defined as~\cite%
{helstrom,wang,yang}
\begin{equation}
F_{Q}(\rho (\theta ))=\mathtt{Tr}[\rho (\theta )L^{2}].
\end{equation}%
The symmetric logarithmic derivative operator $L$ is determined by%
\begin{equation}
\partial _{\theta }\rho (\theta )=\frac{1}{2}[L\rho (\theta )+\rho (\theta
)L].
\end{equation}%
Assume that the spectral decomposition of the density operator is given by $%
\rho (\theta )=\sum_{i=1}^{s}p_{i}|\varphi _{i}\rangle \langle \varphi _{i}|$
with eigenvalues $p_{i}$ and eigenvectors $|\varphi _{i}\rangle $ of $\rho
(\theta )$. \ And $L$ can be solved by rewriting the above equation under
the eigenbasis of $\rho (\theta )$. Then the QFI obtained can be written as
\begin{eqnarray}
F_{Q}(\rho (\theta )) &=&4\sum_{i=1}^{s}p_{i}(\langle \partial _{\theta
}\varphi _{i}|\partial _{\theta }\varphi _{i}\rangle -|\langle \varphi
_{i}|\partial _{\theta }\varphi _{i}\rangle |^{2})  \nonumber  \label{QFI} \\
&&-\sum_{i\neq j}^{s}\frac{8p_{i}p_{j}}{p_{i}+p_{j}}|\langle \varphi
_{i}|\partial _{\theta }\varphi _{j}\rangle |^{2}.
\end{eqnarray}%
The value of parameter $\theta $ can be estimated through measuring $\rho
(\theta )$. From the QFI, we obtain the lower bound of the variance of the
estimator for the parameter $\theta $, given by the quantum Cram\'{e}r-Rao
(QCR) theorem:~\cite{helstrom,holevo} $(\delta\theta )^{2}\geq 1/F_{Q}(\rho
(\theta ))$. From this inequality, it is obvious that the variance of the
estimation is small for a large value of $F_{Q}(\rho (\theta ))$.

The estimation of parameters in general quantum metrology process consists
of the following steps. First, we prepare a quantum state $\rho _{in}$. Then
the system undergoes the $\theta$-dependent process $e^{-i\theta U}$ with
the phase-shift generator $U$, and evolves to the state $\rho (\theta
)=e^{-i\theta U}\rho _{in}e^{i\theta U}$. Finally, we estimate the parameter
$\theta $ by measuring $\rho (\theta )$. The variance of the estimation is
bounded by the inverse of the QFI. As the eigenvalues of $\rho (\theta )$
and $\rho _{in}$ are the same, the expression of QFI (~\ref{QFI}) is
simplified as
\begin{equation}
F_{Q}(\rho (\theta ),U)=4\sum_{i=1}^{s}p_{i}(\delta U)^{2}-\sum_{i\neq j}^{s}%
\frac{8p_{i}p_{j}}{p_{i}+p_{j}}|\langle \varphi _{i}|U|\varphi _{j}\rangle
|^{2}.  \label{QFI1}
\end{equation}%
where $(\delta U)^{2}=\langle \varphi _{i}|U^{2}|\varphi _{i}\rangle
-|\langle \varphi _{i}|U|\varphi _{i}\rangle |^{2}$. For a pure state $\rho
_{in}=|\varphi \rangle \langle \varphi |$, and the QFI is simplified as $%
F_{Q}(\rho (\theta ),\widehat{U})=4\sum_{i=1}^{s}p_{i}(\delta U)^{2}$. It is
obvious that the QFI depends on the quantum state $\rho _{in}$ and the
choice of the phase-shift generator $U$. In our work, since the Dicke model is
the collective atom model, the quantum state $\rho _{in}$ is prepared in a
specific class. For collective models, all two-level atoms are completely
equivalent and the ground states of collective models are invariant under
the permutation group. We discuss the QFI of the reduced two-atom state
which is a symmetric state and permutation-invariant in the Dicke model. The
QFI is expected to capture the quantum criticality and signal the presence
of QPT. Moreover, it is significant to compute the finite-size scaling
exponent of the QFI for the universality of the QPT.

\section{the QFI of the reduced two-atom state}

We study the QFI and its scaling behavior for $N$ two-level atoms system in
the Dicke model. The Dicke Hamiltonian can be written in terms of the
collective momentum form~\cite{Emary,chen}
\[
~H={\omega }a^{\dag }a+{\Delta }J_{z}+\frac{2{\lambda }}{\sqrt{N}}(a^{\dag
}+a)J_{x},
\]%
where $a^{\dag }$ and $a$ are the bosonic annihilation and creation
operators of the single-mode cavity, $\Delta $ and $\omega $ are the
transition frequency of the two-level atom and the frequency of the single bosonic
mode, $\lambda $ is the coupling constant. $J_{x}$ and $J_{z}$ are the
collective atomic operators as $J_{\alpha }=\frac{1}{2}\sum_{i=1}^{N}\sigma
_{i,\alpha }$ ($\alpha =x,y,z$). And the Hilbert space of this algebra is
spanned by the Dicke state $\{|j,m\rangle ,m=-j,-j+1,...,j-1,j\}$ with $%
j=N/2 $, which is the eigenstate of $J^{2}$ and $J_{z}$.

The Hilbert space of the total system can be expressed in terms of the basis
$\{|n\rangle \otimes |j,m\rangle \}$, where $|n\rangle $ is the Fock state
containing $n$ bosons. To our best knowledge, it is very difficult to obtain
convergent results for large number of atoms based on usual basis of the
Fock states~\cite{Hirsch,lambert1}. An extended coherent state technique~%
\cite{chen} has been proposed to the Dicke model up to large $N$-atom
system, which has been confirmed recently by comparing with the results in
terms of basis of the Fock states~\cite{Hirsch}. In the extended coherent
state approach, the wave function can be expressed in terms of the basis $%
\{\left\vert \varphi _{m}\right\rangle _{b}\otimes \left\vert
j,m\right\rangle \}$, where $\left\vert \varphi _{m}\right\rangle _{b}$ is
the bosonic extended coherent state
\begin{equation}
\left\vert \varphi _{m}\right\rangle _{b}=\sum_{k=0}^{N_{tr}}c_{m,k}\frac{1}{%
\sqrt{k!}}(a^{\dag }+g_{m})^{k}e^{-g_{m}a^{\dag }-g_{m}^{2}/2}\left\vert
0\right\rangle _{a},  \label{wavefunction}
\end{equation}%
where $g_{m}=2m\lambda/(\omega \sqrt{N})$, $N_{tr}$ is the truncated bosonic
number in the space of the new operator $A_{m}=a+g_{m}$, $|0{\rangle }_{a}$
is the vacuum as $a|0{\rangle }_{a}=0$, and the coefficient $c_{m,k}$ can be
determined through the exact diagonalization. Then the ground state for the
finite-$N$ system takes the form of
\begin{equation}
|G\rangle =\sum_{m=-j}^{j}\left\vert \varphi _{m}\right\rangle _{b}\otimes
|j,m\rangle .  \label{gs}
\end{equation}%
In the ground state it undergoes a transition from the normal to the
super-radiant phase when increasing the coupling $\lambda$ throught a
critical value of $\lambda _{c}=\sqrt{{\omega }{\Delta }}/2$~\cite%
{lambert,chen,emary1,emary2,benedict,Emary}. In the super-radiant phase, the
atomic ensemble spontaneously emits with an intensity proportional to $N^2$
rather than $N$.

Since there is on going interest in studying the connection between the QFI and QPT.
The QFI is expected to shed light on the dramatic effects occurring at the
critical point, providing to signal the presence of QPT. It is necessary to
study the QFI in the ground state of the Dicke model.

In the quantum metrology process, we choose the reduced two-atom state $\rho$
associated with arbitrary two atoms as a prepared state, which is obtained
by tracing out the density matrix over all other $N-2$ atoms and field mode.
For collective atom ensemble, the reduced two-atom state is particularly
well suited since it does not depend on the two atoms selected, all atoms
being completely equivalent. Since the $N$-atom state is the Dicke state $|j,m>$,
which is a symmetric state. The reduced two-atom state can be extracted from
the symmetric multi-atom Dicke state $|j,m>$. Due to the symmetry of the state of $N$-atoms
under exchange of atoms, the reduced two-atom state $\rho$ is invariant
under the permutation Group. The matrix elements of the reduced two-atom state $\rho$
can be expressed in terms of the averages of the collective atomic
operators, which has been addressed by Wang and J.Vidal in Ref.~\cite%
{wang1,julien1}. Then the reduced two-atom state can be given in the basis
of two atoms $\{|{\downarrow }{\downarrow }{\rangle },|{\downarrow }{%
\uparrow }{\rangle },|{\uparrow }{\downarrow }{\rangle },|{\uparrow }{%
\uparrow }{\rangle }\}$ (with ${\sigma _{z}}|{\uparrow }{\rangle }=|{%
\uparrow }{\rangle }$ and ${\sigma _{z}}|{\downarrow }{\rangle }=-|{%
\downarrow }{\rangle }$) as
\begin{equation}
~\rho =\left(
\begin{array}{llll}
v_{+} & x_{+}^{\ast } & x_{+}^{\ast } & u^{\ast } \\
x_{+} & w & y & x_{-}^{\ast } \\
x_{+} & y & w & x_{-}^{\ast } \\
u & x_{-} & x_{-} & v_{-}%
\end{array}%
\right) ,  \label{rho:}
\end{equation}%
where the matrix elements can be represented by the expected values of the
collective spin operators
\begin{eqnarray}  \label{elements}
~v_{\pm } &=&\frac{N^{2}-2N+4{\langle }J_{z}^{2}{\rangle }{\pm }4(N-1){%
\langle }J_{z}{\rangle }}{4N(N-1)},  \label{rho:2} \\
x_{\pm } &=&\frac{(N-1){\langle }J_{+}{\rangle }{\pm }{\langle }%
[J_{+},J_{z}]_{+}{\rangle }}{2N(N-1)},  \nonumber \\
w &=&\frac{N^{2}-4{\langle }J_{z}^{2}{\rangle }}{4N(N-1)},y=\frac{{\langle }%
J_{x}^{2}+J_{y}^{2}{\rangle }-N/2}{N(N-1)},  \nonumber \\
u &=&\frac{{\langle }J_{+}^{2}{\rangle }}{N(N-1)},  \nonumber
\end{eqnarray}%
where $[A,B]_{+}=AB+BA$, and $w=y$ for $\sum_{\alpha =x,y,z}J_{\alpha
}^{2}=J^{2}=\frac{N}{2}(\frac{N}{2}+1)$. $\langle J_{\alpha }\rangle$ and $%
\langle J_{\alpha }^2\rangle (\alpha=x,y,z,+)$ mean the averages of the collective
atomic operators over the
ground state $|G\rangle$, Eq.(~\ref{gs}). Thus, we can calculate the
expected values of the collective atomic operators to determine the elements
of the reduced two-atom state.

Since there is a conserved parity $\Pi$ in the Dicke model, such that $%
[H,\Pi]=0$, which is given by $\Pi=e^{i\pi \hat{N}}$ with the excitation
number $\hat{N}=a^\dagger a+J_z+N/2$. The parity $\Pi$ possesses two
eigenvalues $\pm1$, depending on whether the number of quanta is even or
odd. Then the Hilbert space of the total system is split into two
noninteracting subspaces, resulting $\langle G|J_{\pm }|G\rangle=0$. Then it
is easily to find $x_{\pm }=0$ in Eq.(~\ref{rho:2}). Hence the reduced
two-atom state can be shown in $X$ form as
\begin{equation}
~\rho =\left(
\begin{array}{llll}
v_{+} & 0 & 0 & u^{\ast } \\
0 & w & y & 0 \\
0 & y & w & 0 \\
u & 0 & 0 & v_{-}%
\end{array}%
\right) .  \label{rho:1}
\end{equation}%
The reduced two-atom state facilitates the analytical evaluation of the QFI
of the two-atom state in the following.

We consider an estimation of the parameter $\theta $ introduced by the
following unitary transformation $S=\exp (-i\theta \sigma _{z}^{\theta })$ with the
phase-shift generator $\sigma _{z}^{\theta }=\sigma _{z}\otimes I$~%
\cite{jing}. Here $I$ is the $2\times 2$ identity matrix and $\sigma _{z}$
is the pauli matrix. From the definition in Eq.~(\ref{QFI1}), the analytical
expression of the QFI are obtained as
\begin{eqnarray}
F_{Q}(\rho ,\sigma _{z}^{\theta }) &=&4\sum_{i=1}^{s}p_{i}(\delta \sigma
_{z}^{\theta })^{2}-\sum_{i\neq j}^{s}\frac{8p_{i}p_{j}}{p_{i}+p_{j}}%
|\langle \varphi _{i}|\sigma _{z}^{\theta }|\varphi _{j}\rangle |^{2}
\nonumber  \label{XQFI} \\
&=&16(\frac{u^{2}}{v_{+}+v_{-}}+\frac{w}{2})
\end{eqnarray}%
which are evaluated in detail in Appendix. We calculate the QFI in the
infinite and finite size systems as $F_{Q,\infty}$ and $F_{Q,N}$
respectively .

We begin to discuss the QFI of the reduced two-atom state in $X$ form by
evaluating the expected values of matrix elements in the thermodynamics
limit, in which the number of atoms becomes infinite. In this limit
analysis, we first apply the Holstein-Primakoff transformation to change the
collective angular operators to the boson operators $b(b^{\dag })$ by $%
J_{+}=b^{\dag }\sqrt{N-b^{\dag }b}$, $J_{-}=\sqrt{N-b^{\dag }b}b$, and $%
J_{z}=b^{\dag }b-N/2$, where $[b,b^{\dag }]=1$~\cite{Emary}. Then the
displacements of the boson operators are introduced to depict the behaviors
of super-radiation phase as $c^{\dag }=a^{\dag }+\sqrt{N}{\alpha }$ and $%
d^{\dag }=b^{\dag }-\sqrt{N}{\beta }$. By means of the boson expansion
approach, we expand the Hamiltonian with respect to the new operator $%
c^{\dag }$ and $d^{\dag }$ as power series in $1/N,$
\begin{equation}
~H=NH_{0}+N^{1/2}H_{1}+\cdots ,
\end{equation}%
where $H_{0}={\omega \alpha }^{2}-4{\lambda \alpha \beta }\sqrt{1-{\beta }%
^{2}}+{\Delta (\beta }^{2}-1/2)$ and $H_{1}=-{\alpha \omega +2{\lambda }%
\beta (1-{\beta }^{2}/2)(c^{\dag }+c)+\Delta \beta }(d^{\dag }+d)$. By using
large $N$ expansions of $H$ up to the $1/N$, we obtain the ground state
energy $E_{G}(\alpha ,\beta )$ as
\begin{equation}
\frac{E_{G}(\alpha ,\beta )}{N}={\omega }{\alpha }^{2}-4{\lambda }{\alpha }{%
\beta }\sqrt{1-{\beta }^{2}}+{\Delta }({\beta }^{2}-\frac{1}{2}).
\end{equation}%
Minimizing the ground state energy gives
\begin{eqnarray}
{\omega }{\alpha }-2{\lambda }{\beta }\sqrt{1-{\beta }^{2}} &=&0 \\
2{\alpha }{\lambda }\sqrt{1-{\beta }^{2}}-\frac{2{\alpha }{\lambda }{\beta }%
^{2}}{\sqrt{1-{\beta }}}-{\beta }{\Delta } &=&0.  \nonumber
\end{eqnarray}%
then we have
\begin{eqnarray}
\beta ^{2} &=&\max \{0,\frac{1}{2}(1-{\mu })\}, \\
\alpha &=&\frac{2\lambda }{\omega }{\beta }\sqrt{1-\beta ^{2}},  \nonumber
\end{eqnarray}%
where $\mu =1$ in the normal phase and $\mu =(\lambda _{c}/\lambda )^{2}$ in
the superradiant phase with the critical point $\lambda _{c}=\sqrt{{\omega }{%
\Delta }}/2$. Next we can derive the matrix elements of the reduced two-atom
state $\rho$ in Eq.~(\ref{rho:1}) up to $O(1)$
\begin{eqnarray}
v_{+} &=&\beta ^{4},~v_{-}=(1-{\beta }^{2})^{2},  \label{parameter:1} \\
w &=&y={\beta }^{2}(1-{\beta }^{2}),  \nonumber \\
~u &=&{\beta }^{2}(1-{\beta }^{2}).  \nonumber
\end{eqnarray}%
In the thermodynamics limit $N\rightarrow \infty $, the QFI of the reduced
two-atom state $F_{Q,\infty }$ can be expressed as%
\begin{equation}
F_{Q,\infty }=\frac{8{\beta }^{2}(1-{\beta }^{2})}{\beta ^{4}+(1-{\beta }%
^{2})^{2}}.  \label{XQFI3}
\end{equation}%
We know that $\beta ^{4}+(1-{\beta }^{2})^{2}\geq 2\beta ^{2}(1-\beta ^{2})$%
, then one can find that the maximum QFI is $F_{Q,\infty }^{max}=4$, which
gives the minimum value of variance of the estimator $\delta \theta =1/2$.

In the thermodynamics limit, there is no excitation of the system in the
normal phase, and the atoms part of the ground state is the pure Dicke state
$|j,-j\rangle =\prod_{k=1}^{N}|\downarrow \rangle _{k}$, with $\sigma
_{k}^{z}|\downarrow \rangle _{k}=-|\downarrow \rangle _{k}$. Then the QFI of
the reduced two-atom state in Eq.(~\ref{XQFI}) is simplified by $F_{Q}(\rho
,\sigma _{z}^{\theta })=4(\delta \sigma _{z}^{\theta })^{2}$. For the ground
state $|j,-j\rangle , $ the variance of $(\delta \sigma _{z}^{\theta })^{2}$
equals to $0$ and hence one obtain the minimum value of the QFI $F_{Q,\infty
}^{min}=0$. The minimum value leads to the maximum variance of the estimator
$\delta \theta \rightarrow \infty $ in the normal phase. It also can be
easily obtained $F_{Q,\infty }=0$ from Eq.(~\ref{XQFI3}) with $\beta =0$. In
the superradiant phase for a large coupling strength $\lambda \gg \lambda
_{c}$ with $\beta \rightarrow 1/2$, the QFI of the reduced two-atom state
approaches to the maximum value $F_{Q,\infty }\rightarrow 4$. In the strong
coupling limit, the ground state (~\ref{gs}) in the $J_{x}$-representation
can be described by $|G\rangle =\frac{1}{\sqrt{2}}[(\prod%
\limits_{k=1}^{N}|e_{x}\rangle _{k})|0\rangle
_{A_{-N/2}}+(\prod\limits_{k=1}^{N}|g_{x}\rangle _{k})|0\rangle _{A_{N/2}}]$%
, where $\sigma _{k}^{x}|e_{x}\rangle _{k}=|e_{x}\rangle _{k}$ and $\sigma
_{k}^{z}|g_{x}\rangle _{k}=-|g_{x}\rangle _{k}$. The vacuum coherent states $%
|0\rangle _{A_{\pm N/2}}$ satisfy $A_{\pm N/2}|0\rangle _{A_{\pm N/2}}=0$.
It yields the variance of $(\delta \sigma _{z}^{\theta })^{2}=1$ and the QFI
$F_{Q,\infty } $ equals to $4$. It indicates that the variance of the
estimator $\delta \theta$ is minimum in the ground state in the superradiant
phase, which can be distinguished from that in the normal phase.

\begin{figure}[tbp]
\includegraphics[scale=0.5]{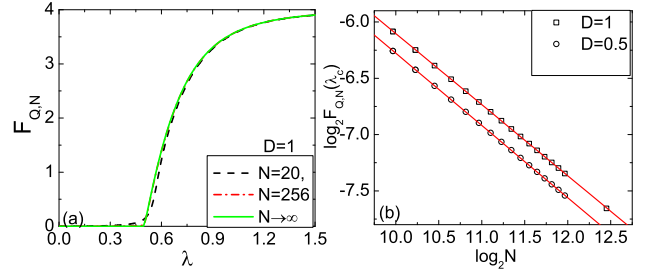}
\caption{(Color online) QFI $F_{Q,N}$ of the reduced two-atom state in the
Dicke model as a function of the coupling constant $\protect\lambda$ with
different sizes $N=20,256$ and $N\rightarrow\infty$ for $D=1$. (b) Scaling
of $F_{Q,N}$ as a function of $N$ on a log-log scale at the critical point $%
\protect\lambda_c$ for $D=0.5$ and $1$, and the solid line scales as $%
N^{-0.65\pm0.01}$.}
\label{figure1}
\end{figure}

Fig.(~\ref{figure1})(a) displays the QFI of the reduced two-atom state in
the thermodynamics limit for detunings $D=\Delta /\omega=1$. In the normal
phase the QFI equals to zero. As the the coupling strength enters into the
super-radiant phase $\lambda >\lambda _{c}$, the QFI shows monotonous
increasing behaviors, demonstrating the existence of the QPT at the critical
point $\lambda _{c}=\sqrt{\Delta \omega }/2$. As $\lambda $ approaches $%
\lambda \rightarrow \infty $ limit, the QFI $F_{Q,\infty }$ tends to the
maximum value $4$.

\section{finite-size scaling behavior of the QFI}

To study the universality of the superradiant phase transition, the exact
finite-size scaling law is quite important. We illustrate the scaling
behavior of the QFI of the reduced two-atom state. The finite-size scaling
ansatz for the singular functions $F_{Q,N}^{\sin g}$ of the QFI $F_{Q,N}$ in
the vicinity of the critical point is ~\cite{vidal1}
\begin{equation}
F_{Q,N}^{\sin g}(\lambda )\simeq \frac{(\lambda -\lambda _{c})^{\xi }}{N^{n}}%
f \lbrack N(\lambda -\lambda _{c})^{3/2}],  \label{xscal}
\end{equation}%
where $f$ is a function depending on the scaling variable $N(\lambda
-\lambda _{c})^{3/2}$ and $\xi $, $n$ are exponents. To cure the singularity
coming from $(\lambda -\lambda _{c})^{\xi }$, one has $f(x)\sim x^{-2\xi/3}$%
, which leads to $F_{Q,N}^{\sin g}(\lambda _{c})\sim N^{-(n+2\xi /3)}$.

In finite-size atom system, the QFI of the reduced two-atom state is
calculated by the expected value of the collective atomic operators in Eq.(~%
\ref{rho:2}) in the ground state $|G\rangle$ using the bosonic coherent state
technique. Fig.(~\ref{figure1})(a) shows the behavior of $F_{Q,N}$ for
different system sizes $N=20$ and $256$. In the normal phase, it tends to
zero as the number of atoms $N$ increase. And in the super-radiant phase, $%
\lambda>\lambda_c $, the QFI increases from zero to the maximum value $4$,
which are consistent with those in the thermodynamics limit. For the large
system size $N=256$, there is nearly no deviation of the behavior of the QFI
from that in the thermodynamic limit. It facilitates the calculation of the
scaling behavior at the critical point by the bosonic coherent state
approach.

We plot the QFI $F_{Q,N}$ as a function of $N$ on a log-log scale at the
critical point $\lambda_c$ for different detunings $D=0.5$ and $1$, as shown
in Fig.(~\ref{figure1})(b). It is very interesting to observe a power law
scaling $F_{Q}(\lambda_c)\propto N^{\nu}$. Due to the advantage of the
bosonic coherent state technique~\cite{chen}, we are able to study the atom
number up to $N=4000$ atoms. The asymptotic slop in the log-log scale for
the finite size systems gives a exponent $\nu =-0.65\pm 0.01$.

Moreover, the QFI $F_{Q,N}$ in Eq.(~\ref{XQFI}) can be given explicitly in
terms of the collective atomic operators as
\begin{equation}
F_{Q,N}=\frac{32{\langle }J_{+}^{2}{\rangle }^{2}}{N(N-1)(N^{2}-2N+4{\langle
}J_{z}^{2}{\rangle )}}+\frac{2N^{2}-8{\langle }J_{z}^{2}{\rangle }}{N(N-1)}.
\label{XQFI2}
\end{equation}%
Since the finite-size scaling exponents of the collective spin operators in
the Dicke model have been derived by Vidal and Dusuel~\cite{vidal1}, such as
$\langle J_{z}^{2}\rangle /N^{2}\sim N^{-2/3}$ , $\langle J_{y}^{2}\rangle
/N^{2}\sim N^{-4/3}$ and $\langle J_{x}^{2}\rangle /N^{2}\sim N^{-2/3}$.
Thus the finite-size scaling exponent of the QFI in Eq.(~\ref{XQFI2}) can be
directly checked. From those, it is easily to obtain the finite-size scaling
behavior of QFI exactly as $F_{Q,N}\sim N^{-2/3}$. The result is in consistent
with numerical accurate exponent. To the best of our knowledge, such a
finite size scaling for the QFI itself has never been reported in Dicke
model.

In general quantum metrology process, the QFI depends on the prepared
quantum state and the choice of the unitary operator. The QFI of the reduced
$N$-atom state $\rho _{A}$, which is the reduced atomic density matrix by
tracing over the field degree of the freedom, has been studied in small-size
systems $N=20$ in the Dicke model~\cite{jin}. It is very difficult to
predict the finite-size scaling exponent due to the too small system sizes
investigated based on the basis of the Fock states. Here we focus on the
exponent of the QFI of the reduced atomic subsystem for a large number of
atoms based on the ground state in Eq.(~\ref{gs}). The QFI $F_{A}$ of the
reduced atomic state $\rho _{A}(\theta )=e^{-i\theta J_{z}}\rho
_{A}e^{i\theta J_{z}}$ can be given by Eq.(~\ref{QFI1}), where the phase-shift
generator $U$ is replaced by $J_{z}$ and {$|\varphi _{i}\rangle$ are
the corresponding eigenvectors of $\rho _{A}$ with nonzero eigenvalues $%
\{p_{i}\}$. In the thermodynamics limit, the analytical results of the
scaled QFI for the atomic subsystem is~\cite{jin}
\begin{equation}
F_{A,\infty }/N=\frac{2\mu \Delta }{\varepsilon _{+}+\varepsilon
_{-}+(\Delta ^{2}/\mu ^{2}-\omega ^{2})/(\varepsilon _{+}+\varepsilon _{-})},
\end{equation}%
where the excitation energies is given by $\varepsilon _{\pm }^{2}=\frac{1}{2%
}(\omega ^{2}+\Delta ^{2}/\mu ^{2})\pm \frac{1}{2}\sqrt{(\omega ^{2}-\Delta
^{2}/\mu ^{2})^{2}+16\lambda ^{2}\omega \Delta \mu }$. As the coupling
strength $\lambda $ approaches to the critical point as $\lambda \rightarrow
\lambda _{c}$, the excitation energy of $\varepsilon _{+}$ tends towards a
value of $\sqrt{\omega ^{2}+\Delta ^{2}}$ and $\varepsilon _{-}$ vanishes, $%
\varepsilon _{-}\rightarrow 0$. The critical exponents of the QPT can be
manifested in the behavior of the scaled QFI
\begin{equation}
\frac{F_{A,\infty}}{N}(\lambda \rightarrow \lambda _{c})\simeq \frac{\sqrt{\omega
^{2}+\Delta ^{2}}}{\Delta }+\sqrt{\frac{32\omega ^{2}\lambda _{c}^{3}}{%
\Delta ^{2}(16\lambda _{c}^{4}+\omega ^{4})}}|\lambda _{c}-\lambda |^{1/2}.
\label{atomscal}
\end{equation}%
Fig.(~\ref{figure2})(a) displays the maximum value $\sqrt{\omega ^{2}+\Delta
^{2}}/\Delta$ of the scaled QFI $F_{A,\infty }/N$ at the critical point $%
\lambda _{c}$. As addressed in Ref.~\cite{jin}, the scaled QFI is larger
than $1$ in the normal phase, and then decreases to zero in the superradiant
phase.

It is interesting to observe the finite-size scaling behavior of the scaled QFI $F_{A}/N$ in
the Dicke model. We calculate $F_{A}/N$ by the ground state $|G\rangle$ in Eq.(~\ref{gs}) using the
bosonic coherent-state technique. Fig.(~\ref{figure2})(a) displays the scaled QFI $F_{A}/N$
for finite atom ensemble $N=20$ and $N=256$.  The scaling behavior of $(F_{A,\infty }-F_{A})/N$ as
a function of $N$ at the critical point $\lambda _{c}$ is shown in Fig.(~\ref{figure2})(b) for different values
of $D=0.5$ and $1$ on a log-log scale.
A power-law behavior exists at large $N$. One can see that the finite-size
exponents extracted from all curves tend to a converging value $-0.33\pm
0.01$ in Fig.(~\ref{figure2})(b). The precise estimate of the exponent for
the QFI is very significant to help clarify the universality of the QPT.
\begin{figure}[tbp]
\includegraphics[scale=0.5]{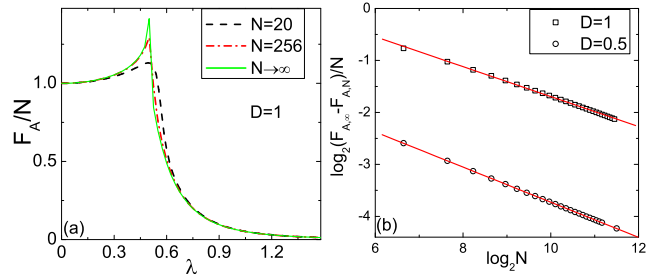}
\caption{(Color online) (a)The scaled QFI $F_{A}/N$ of the reduced $N$-atom
state as a function of $\protect\lambda$ with different sizes $N=20$, $256$
and $N\rightarrow\infty$ for $D=1$. (b)Scaling of the scaled QFI $%
(F_{A,\infty }-F_{A})/N$ as a function of $N$ on a log-log scale at the
critical point $\protect\lambda_c$ for $D=0.5$ and $1$, and the solid line
scales as $N^{-0.33\pm0.01}$.}
\label{figure2}
\end{figure}

\section{Conclusion}

In summary, we have proposed the QFI of the reduced two-atom state
to capture the quantum criticality of the QPT in the Dicke model from
the quantum estimation perspective. The QFI are obtained analytically in the
thermodynamics limit. The behavior of the QFI of the reduced two-atom state shows
that the ground state undergoes a
superradiant phase transition at the critical regime in the infinite size system. And Finite-size
scaling exponents of the QFI are calculated up to a large atom number $N=4000$
by the bosonic coherent technique. Power law scaling behavior at the
critical point is observed. Moreover, large-$N$ scaling behavior of the scaled QFI of
the reduced $N$-atom state is also calculated, giving the accurate exponent.
Such a scaling behavior of QFI has not been reported in the critical systems
of Dicke model, as far as we know. These salient features might be used for
quantum metrology and quantum estimation in some experimentally realized
systems to the quantum information science and the quantum computing.

\section{Acknowledgements}

We thank Julien Vidal for helpful discussion. This work was supported by
National Natural Science Foundation of China (Grants No.~11104363 and
No.11274403), and Research Fund for the Central Universities (No.
CQDXWL-2013-Z014).

\section*{Appendix}

For the reduced two-atom state $\rho$ in Eq.(~\ref{rho:1}), the
corresponding eigenvalues are given by%
\begin{equation}
p_{1}=2w,p_{2}=0,p_{\pm }=\frac{1}{2}(v_{+}+v_{-}\pm\sqrt{\gamma }),
\end{equation}%
where $\gamma =(v_{+}-v_{-})^{2}+4|u|^{2}$. The corresponding eigenstates to
$p_{1}$ and $p_{\pm }$ are%
\begin{eqnarray}
|\phi _{1}\rangle &=&\frac{1}{\sqrt{2}}\left(
\begin{array}{c}
0 \\
1 \\
1 \\
0%
\end{array}%
\right) , \\
|\phi _{\pm }\rangle &=&\epsilon _{\pm }\left(
\begin{array}{c}
(v_{+}-v_{-}\pm \sqrt{\gamma })/2u \\
0 \\
0 \\
1%
\end{array}%
\right) ,
\end{eqnarray}%
with $\epsilon _{\pm }^{2}=2u/\sqrt{\gamma \pm (v_{+}-v_{-})\sqrt{\gamma }}$%
. For the unitary operation $S=\exp (-i\theta \sigma _{z}^{\theta })$, the
QFI of the reduced two-atom state can be evaluated as
\begin{eqnarray}
F_{Q} &=&4p_{\pm }\langle \delta \sigma _{z}^{\theta }\rangle _{\pm
}^{2}+4p_{1}\langle \delta \sigma _{z}^{\theta }\rangle _{1}^{2}-\frac{%
16p_{+}p_{-}}{p_{+}+p_{-}}|\langle \phi _{+}|\sigma _{z}^{\theta }|\phi
_{-}\rangle |^{2}  \nonumber \\
&&-\sum_{i=\pm }\frac{16p_{i}p_{1}}{p_{i}+p_{1}}|\langle \phi _{i}|\sigma
_{z}^{\theta }|\phi _{1}\rangle |^{2},
\end{eqnarray}%
where the variance of operator $\sigma _{z}^{\theta }$ is $\langle \delta
\sigma _{z}^{\theta }\rangle _{i}^{2}=\langle \phi _{i}|(\sigma _{z}^{\theta
})^{2}|\phi _{i}\rangle -|\langle \phi _{i}|\sigma _{z}^{\theta }|\phi
_{i}\rangle |^{2}$. Substituting the values of $p_{\pm ,1}$ and $|\phi _{\pm
,1}\rangle $ into the above equation, the QFI can be given analytically.

*Corresponding author. Email:yuyuzh@cqu.edu.cn.

\end{document}